\documentclass[runningheads,a4paper]{llncs}

\usepackage{amssymb}
\usepackage{graphicx}

\usepackage{url}
\urldef{\mailsa}\path|{mark.scanlon, tahar.kechadi}@ucd.ie|    
\newcommand{\keywords}[1]{\par\addvspace\baselineskip
\noindent\keywordname\enspace\ignorespaces#1}

\begin{document}

\mainmatter  

\title{Digital Evidence Bag Selection for \\P2P Network Investigation}

\titlerunning{Digital Evidence Bag Selection for P2P Network Investigation}

%
%
\author{Mark Scanlon\and Tahar Kechadi}
\authorrunning{Mark Scanlon\and Tahar Kechadi}

\institute{School of Computer Science and Informatics,\\
University College Dublin, \\
Belfield, Dublin 4, Ireland.\\
\mailsa\\
\url{http://csi.ucd.ie}}

%
%

\toctitle{Digital Evidence Bag Selection for P2P Network Investigation}

\tocauthor{Mark Scanlon and Tahar Kechadi}
\maketitle

\begin{abstract}
The collection and handling of court admissible evidence is a fundamental component of any digital forensic investigation. While the procedures for handling digital evidence take much of their influence from the established policies for the collection of physical evidence, due to the obvious differences in dealing with non-physical evidence, a number of extra policies and procedures are required. This paper compares and contrasts some of the existing digital evidence formats or ``bags'' and analyses them for their compatibility with evidence gathered from a network source. A new digital extended evidence bag is proposed to specifically deal with evidence gathered from P2P networks, incorporating the network byte stream and on-the-fly metadata generation to aid in expedited identification and analysis.
\end{abstract}

\section{Introduction}

Selecting an evidence storage format for any digital investigation can be an important decision with respect to the efficiency of the entire investigation. The are numerous choices of digital evidence bags (DEBs) available with varying levels of inter-compatibility and performance benefits. Most of the existing DEBs are focused on the storage of images from physical digital storage media, e.g., hard drives, memory cards, etc. However, with respect to network focused investigations, many of the existing procedures and tools merely offer a method for capturing the network traffic in a ``raw'' format. The analysis stage of the investigation is generally limited to start after the completion of the network recording/sniffing process. In a non-time-sensitive network investigation, this limitation might have a minor impact on the desired evidence collection. However, in a time-sensitive, high traffic network, this limitation could result in the analysis, and possible reaction/mitigation phase of the network investigation commencing too late, e.g., in a high packet frequency P2P network such as a P2P botnet. Due to the high churn rates typical of most P2P networks, the time required from the recording of P2P network traffic to identifying the traffic and extracting any useful evidence is paramount. 

When attempting to reverse engineer a newly discovered P2P network, the job is left in the hands of the digital investigator. It is his responsibility to attempt to match common packet patterns and deduce based on the gathered evidence the protocol of the network. This paper outlines the case for a P2P network specific DEB which is capable of storing the captured packets alongside some useful metadata to aid the investigator in this process.

\section{Evidence Handling}
\label{handling}

When dealing with physical forensic evidence, the commonly used handling procedure is the ``chain of custody" \cite{debq}. The chain of custody commences at the crime scene where the evidence is collected, when the investigating officer collects any evidence he finds and places it into an evidence bag. This evidence bag will be sealed to avoid any contamination from external sources and signed by the officer and will detail some facts about the evidence, e.g., description of evidence, location it was found, date and time found, etc. The chain of custody will then be updated again when the evidence is checked into the evidence store in the forensic laboratory. When it comes to analysing the evidence, it will be checked out to the forensic analyst's custody and any modification to the evidence required to facilitate the investigation, e.g., taking a sample from a collected fibre to determine its origin or identification, etc., will be logged and documented. 

The procedures outlined above for physical evidence need to be extended for digital evidence acquisition and analysis. Due to the fact that digital evidence is generally analysed on forensic workstations, most of the above sequences can be automated into concise logging of all interactions. During a digital investigation, there is no requirement to modify the existing evidence in any way. This is because all analysis is conducted on an image of the original source and any discovered evidence can be extracted from this image, documented and stored separately to both the original source and the copied image. It is imperative when dealing with all types of evidence that all procedures used are reliable, reproducible and verifiable. In order for evidence to be court admissible, it must pass the legal criteria for the locality that the court case is being heard, as outlined in greater detail in section \ref{ch2:legal} below.


\subsection{What does ``Forensically Sound" really mean?}
\label{ch2:sound}

Many of the specifications for digital forensic acquisition tools, analysis tools, storage formats and hash functions state that the product in question is ``forensically sound" or that the product works with the digital evidence in a ``forensically sound manner", without specifying exactly what the term means. In 2007, E. Casey published a paper in the Digital Investigation Journal entitled ``What does ``forensically sound" really mean ?" \cite{forensicallysound}. 

In this paper, Casey outlines some of the common views of forensic professionals regarding dealing with digital forensic evidence. Purists state that any digital forensic tools should not alter the original evidence in any way. Others point out that the act of preserving certain types of evidence necessarily alters the original, e.g., a live memory evidence acquisition tool must be loaded into memory (altering the state of the volatile memory and possibly overwriting some latent evidence) in order to run the tool and capture any evidence contained in the memory. Casey then goes onto to explain how some traditional forensic process require the altering of some of the evidence in order to collect the required information. For example, collecting DNA evidence requires taking a sample from some collected evidence, e.g., a hair. Subsequently, the forensic analysis of this evidentiary sample (DNA profiling) is destructive in its nature which further alters the original evidence.


\subsection{Splitting Evidence}
\label{ch2:splitting}

It is not always possible to store the entire image of a particular storage device or sequence of network packets in one large file. This could be for a number of reasons, such as the evidence being stored on a FAT32 formatted hard drive which is only capable of addressing a file less than $2^{32}$ bytes (4,294,967,296 bytes or 4 gigabytes) or if evidence needs to be backed up to external media, e.g., a data CD or DVD, capable of storing 700MB and 4.7GB respectively. If the acquired evidence is going to be transmitted over the Internet, it should be a capability of any digital evidence bag to split the evidence into smaller parts to minimise the cost of dropped connections. The CDESF working group conducted a survey in 2006 and found that each of the evidence storage formats they tested was capable of allowing split archiving and storage of evidence \cite{dfrws2006}. 

Should any tool split the evidence during acquisition, for transmission or storage purposes, this collected evidence should be recompilable into the original source for examination purposes. To ensure forensic integrity, the tools used for splitting and recompiling the evidence should be able to verify the recompiled image against the original untouched source using a sufficiently collision resistant hash values.

\section{Evidence Storage Formats}
\label{storage}

There is currently no universal standard for the format that digital evidence and any case related information is stored. This is due to the fact that there are no state or international governmental policies to outline a universal format. Many of the vendors developing forensic tools have developed their own proprietary format. With such a relatively small target market, it sometimes makes business sense for them to try and lock their customers into buying only their software in the future. There have been a small number of attempts at creating open formats to store evidence and any related metadata. This section describes the most common of these formats below.

\subsection{Common Digital Evidence Storage Format}
\label{ch2:cdesf}

The Common Digital Evidence Storage Format (CDESF) Working Group was created as part of the Digital Forensic Research Workshop (DRFWS) in 2006. The goal of this group was to create an open data format for storing digital forensic evidence and associated metadata from multiple sources, e.g., computer hard drives, mobile Internet devices, etc. \cite{cdesf}. The format which the CDESF working group were attempting to create would have specified metadata capable of storing case-specific information such as case number, digital photographs of any physical evidence collected and the name of the digital investigator conducting the investigation. In 2006, the working group produced a paper outlining the advantages and disadvantages of various evidence storage formats \cite{dfrws2006}.

Ultimately due to resource restrictions, the CDESF working group was disbanded in 2007 before accomplishing their initial goal.

\subsection{Raw Format}
\label{ch2:raw}

According to the CDESF Working Group, ``the current de facto standard for storing information copied from a disk drive or memory stick is the so-called ``raw" format: a sector-by-sector copy of the data on the device to a file" \cite{evidencestandards}. The raw format is so-called due to the fact that it is simply a file containing the exact sector-by-sector copy of the original evidence, e.g., files, hard disk/flash memory sectors, network packets, etc. Raw files are not compressed in any manner and as a result, any deleted or partially overwritten evidence that may lay in the slackspace of a hard disk is maintained. All of the commercial and open source digital evidence capturing tools available have the capability of creating raw files.

\subsection{Advanced Forensic Format}
\label{ch2:aaf}

The Advanced Forensic Format (AAF) is an open source, extensible format created by S. Garfinkel in Basis Technology in 2006 \cite{aaf}. The AAF format has a major emphasis on efficiency and as a result is partitioned into two layers; the disk representation layer which defines segment name used for storing all data associated with an image and the data storage layer which defines how the image is stored, be it binary or XML\cite{containers}. The format specifies three variants; AFF, AFD and AFM. AFF stores all data and metadata in a single file, AFD stores the data and metadata in multiple small files, and AFM stores the data in a raw format and the metadata is stored in a separate file \cite{containers}.

\subsection{Generic Forensic Zip}
\label{ch2:gfzip}

Generic Forensic Zip (gfzip) is an open source project to create a forensically sound compressed digital evidence format based on AAF \ref{ch2:aaf} \cite{gfzip}. Due to the fact that it is based upon the AAF format, there is limited compatibility between the two in terms of segment based layout. One key advantage that gfzip has over the AAF format is that gfzip seeks to maintain compatibility with the raw format \ref{ch2:raw}. It achieves this by allowing the raw data to be placed first in the compressed image \cite{containers}.

\subsection{Digital Evidence Bag (QinetiQ)}
\label{ch2:debq}

The method for traditional evidence acquisition involves a law enforcement officer collecting any relevant items at the crime scene and storing the evidence in bags and seals. These evidence bags may then be tagged with any relevant case specific information, such as \cite{debq}:
\begin{itemize}
\item Investigating Agency / Police Force
\item Exhibit reference number
\item Property reference number
\item Case/Suspect name
\item Brief description of the item
\item Date and time the item was seized/produced
\item Location of where the item was seized/produced
\item Name of the person that is producing the item as evidence
\item Signature of the person that is producing the item
\item Incident/Crime reference number
\item Laboratory reference number
\end{itemize}

Digital Evidence Bag (DEB) is a digital version of the traditional evidence bag, created by Philip Turner in 2005 \cite{debq}. DEB is based on an adaptation of existing storage formats, with potentially infinite capacity. The data stored in a DEB is stored in multiple files, along with metadata containing the information that would traditionally be written on the outside of an evidence bag. There are currently no tools released that are compatible with the DEB format.

\subsection{Digital Evidence Bag (WetStone Technologies)}
\label{ch2:debwet}

In 2006, C. Hosmer, from WetStone Technologies Inc., published a paper outlining the design of a Digital Evidence Bag (DEB) format for storing digital evidence \cite{debwet}. This format for storing is independent from the Digital Evidence Bag outlined in \ref{ch2:debq}. The format emerged from a research project funded by the U.S. Air Force Research Laboratory. The motivation for this format was similar to the motivation for that described in \ref{ch2:debq}, i.e., to metaphorically mimic the plastic evidence bag used by crime scene investigators to collect physical evidence such as blood, fibres, hairs etc. This format will be released publicly when complete.

\subsection{EnCase Format}
\label{ch2:encaseformat}

The EnCase format for storing digital forensic is proprietary to the evidence analysis tool of the same name. It is by far the most common evidence storage option used by law enforcement and private digital investigation companies for the acquisition of digital evidence from physical storage media \cite{containers}. Because of the proprietary nature of the format, along with the lack of any formal specification from Guidance Software \cite{guidance}, much remains unknown about the format itself. Some competitors to Guidance Software have attempted to reverse engineer the format to provide an element of cross-compatibility with their tools \cite{aaf}. EnCase stores a disk image as a series of unique compressed pages. Each page can be individually retrieved and decompressed in the investigative computer's memory as needed, allowing a somewhat random access to the contents of the image file. The EnCase format also has the ability to store metadata such as a case number and an investigator \cite{aaf}.

\section{Court Admissible Evidence}
\label{ch2:legal}

Since the United States leads the way with the implementation of many standards in relation to evidence handling and the court admissibility of evidence, many other countries look to the procedures outlined by the United States in this area when attempting to create their own legal procedures \cite{commons}. As a result of this, much of the information available regarding the admissibility of digital forensic evidence into court cases is specifically tailored to the Unites States, but will influence law makers across the globe. Carrier \cite{carrier-open} states that in order for evidence to be admissible into a United States legal proceeding, the scientific evidence (a category which digital forensic evidence falls under in the U.S.) must pass the so-called ``Daubert Test" (as outlined below). The reliability of the evidence is determined by the judge in a pre-trail ``Daubert Hearing". The judge's responsibility in the Daubert Hearing is to determine whether the methodologies and techniques used to identify the evidence was sound, and as a result, whether the evidence is reliable.

\subsection{Daubert Test}
\label{ch2:daubert}


The ``Daubert Test" stems from the United States Supreme Court's ruling in the case of Daubert vs. Merrell Dow Pharmaceuticals (1993) \cite{daubert}. The Daubert process outlines four general categories that are used as guidelines by the judge when assessing the procedure(s) followed when handling the evidence during the acquisition, analysis and reporting phases of the investigation, \cite{carrier-open} and \cite{daubert}:

\begin{enumerate}
\item \emph{Testing} -- Can and has the procedure been tested? Testing of any procedure should include testing of the number of false negatives, e.g., if the tool displays filenames in a given directory, then all file names must be shown. It should also incorporate testing of the number of false positives, e.g. if the tool was designed to capture digital evidence, and it reports that it was successful, then all forensic evidence must be exactly copied to the destination. The U.S. National Institute of Standards and Technology (NIST) have a dedicated group working on Computer Forensic Tool Testing (CFTT) \cite{cftt}.
\item \emph{Error Rate} -- Is there a known error rate of the procedure? For example, accessing data on a disk formatted in a documented file format, e.g., FAT32 or ext2, should have a very low error rate, with the only errors involved being programming errors on behalf of the developer. Acquiring evidence from an officially undocumented file format, e.g., NTFS, may result in unknown file access errors occurring, in addition to the potential programming error rate.
\item \emph{Publication} -- Has the procedure been published and subject to peer review? The main condition for evidence admission under the predecessor to the Daubert Test, the Frye Test, was that the procedure was documented in a public place and undergone a peer review process. This condition has been maintained in the Daubert Test \cite{carrier-open}. In the area of digital forensics, there is only one major peer-reviewed journal, the International Journal of Digital Evidence.
\item \emph{Acceptance} -- Is the procedure generally accepted in the relevant scientific community? For this guideline to be assessed, published guidelines are required. Closed source tools have claimed their acceptance by citing the large number of users they have. The developers of these tools do not cite how many of their users are from the scientific community, or how many have the ability to scientifically assess the tool. However, having a tool with a large user base can only prove acceptance of the tool; it cannot prove the acceptance of the undocumented procedure followed when using the tool.
\end{enumerate}


\section{Legal Considerations of Network Forensics}
\label{legal}

Collecting network traffic can pose legal issues. Deploying a packet sniffing or deep packet inspection device, such as those outlined above, can result in the (intentional or incidental) capture of information with privacy and security implications, such as passwords or e-mail content, etc. As privacy has become a greater concern for computer users and organisations, many have become less willing to cooperate or share any information. For example, most ISPs will now require a court order before providing any information related to suspicious activity on their network \cite{Nist:2012:NSP:2331550}. In Europe, continental legal systems operate on the principle of free introduction and free evaluation of evidence and provide that all means of evidence, irrespective of the form they assume, can be admitted into legal proceedings \cite{karyda2007internet}.


\subsection{Jurisdiction}
\label{ch2:jurisdiction}
One aspect of the use of search and seizure warrants in an Internet environment concerns the geographical scope of the warrant issued by a judge or a court authorising the access to the digital data. In the past, the use of computer-generated evidence in court has posed legal difficulties in common law countries, and especially in Australia, Canada, the United Kingdom and the USA. The countries are characterised by an oral and adversarial procedure. Knowledge from secondary sources is regarded as ``hearsay evidence", such as other persons, books, records, etc., and in principle is inadmissible. However, digital evidence has become widely admissible due to several exceptions to this hearsay rule \cite{karyda2007internet}.

\section{Conclusion}

While a number of the digital evidence bags outlined above are capable of storing captured network streams, the DEB itself offers little additional useful features with respect to the investigation of P2P networks. As a result, a new P2P DEB is proposed capable of aiding in the identification and investigation of these networks. This P2P DEB is capable of storing network traffic, alongside specific P2P network metadata, based on on-the-fly analysis during the network capturing process. Building upon existing network packet recording tools, such as libpcap \cite{libpcap}, a additional packet analysis tool can identify known P2P packets as well as record the categorisation and frequency of each type of captured packet. This additional metadata will aid in the identification of new, undiscovered P2P networks.


\bibliographystyle{splncs}
\bibliography{bibfile}

%
%
%
%
%
%
%
\end{document}